\documentclass[runningheads]{llncs}
\usepackage{graphicx}
\usepackage[skip=0pt]{caption}
\usepackage{booktabs}
\usepackage{numprint}
\npthousandsep{,}
\npdecimalsign{.}
\usepackage[inline]{enumitem}
\usepackage{balance}

\usepackage{xcolor}

\usepackage{verbatim}

 \usepackage{multirow}

\usepackage{graphicx} 
\graphicspath{{img/}} 

\usepackage{acronym}
\acrodef{IR}{information retrieval}

\usepackage{hyperref}
\usepackage{url}

\usepackage{mathrsfs}

\usepackage{amsfonts}
\usepackage{amsmath}
\usepackage{amssymb}
\usepackage{amsmath,amsfonts,bm,dsfont}
\usepackage{booktabs}
\usepackage{graphicx}
\usepackage{makecell}
\usepackage{hhline}
\usepackage{xcolor}
\usepackage{colortbl}
\usepackage[normalem]{ulem}
\usepackage{caption}
\usepackage{subcaption}
\usepackage{enumitem}
\usepackage{wrapfig}

\usepackage{cleveref} 

\usepackage{pgfplotstable}
\usepackage{pgfplots}

\definecolor{lblue}{HTML}{A6CEE3}
\definecolor{lgreen}{HTML}{B2DF8A}
\definecolor{lred}{HTML}{FB9A99}
\definecolor{lorange}{HTML}{FDBF6F}
\definecolor{mblue}{HTML}{80B1D3}
\definecolor{mgreen}{HTML}{B3DE69}
\definecolor{mred}{HTML}{FB8072}
\definecolor{morange}{HTML}{FDB462}
\definecolor{blue}{HTML}{1F78B4}
\definecolor{green}{HTML}{33A02C}
\definecolor{red}{HTML}{E31A1C}
\definecolor{orange}{HTML}{FF7F00}
\definecolor{dblue}{HTML}{08519C}
\definecolor{dgreen}{HTML}{006D2C}
\definecolor{dorange}{HTML}{EC7014}

\usepackage{hyperref}

\usepackage[numbers,square,sort&compress]{natbib}

\newcommand{\header}[1]{\vspace{1mm}\noindent\textbf{#1.}}
\newcommand{\subheader}[1]{\vspace{1mm}\textbf{#1.}}

\newcommand{\OurModel}{CLIP-ITA}
\newcommand{\ourtask}{CtI retrieval task}

\AtBeginDocument{%
  \providecommand\BibTeX{{%
    \normalfont B\kern-0.5em{\scshape i\kern-0.25em b}\kern-0.8em\TeX}}}

\newcommand{\shrink}{\vspace*{-2mm}}

\acrodef{CtI}{category-to-image}

\author{
Mariya Hendriksen\inst{1}
\and
Maurits Bleeker\inst{2}
\and
Svitlana Vakulenko\inst{2}
\\
Nanne van Noord\inst{2}
\and
Ernst Kuiper\inst{3}
\and
Maarten de Rijke\inst{2}
\institute{AIRLab,  University of Amsterdam
\and
University of Amsterdam
\and
Bol.com \\
\email{m.hendriksen@uva.nl}
\email{m.j.r.bleeker@uva.nl}
\email{s.vakulenko@uva.nl}
\email{n.j.e.vannoord@uva.nl}
\email{ekuiper@bol.com}
\email{m.derijke@uva.nl}
}
}

\authorrunning{M.  Hendriksen et al.}

\newcommand\sL{\ensuremath{\mathcal{L}}}

\newcommand\bc{\ensuremath{\mathbf{c}}}

\newcommand\bh{\ensuremath{\mathbf{h}}}

\newcommand\bp{\ensuremath{\mathbf{p}}}

\newcommand\bx{\ensuremath{\mathbf{x}}}

\newcommand\BR{\ensuremath{\mathbb{R}}}

\newcommand\R{\ensuremath{\mathbb{R}}} 

\usepackage{color}

\begin{document}

\title{Extending CLIP for Category-to-image Retrieval in E-commerce}
%

\maketitle 

\begin{abstract}
E-commerce provides rich multimodal data that is barely leveraged in practice.  One aspect of this data is a category tree that is being used in search and recommendation.  However,  in practice,  during a user's session there is often a mismatch between a textual and a visual representation of a given category.
Motivated by the problem,  we introduce the task of category-to-image retrieval in e-commerce and propose a model for the task, \OurModel{}. The model leverages information from multiple modalities (textual, visual, and attribute modality) to create product representations.
We explore how adding information from multiple modalities (textual, visual, and attribute modality) impacts the model's performance. In particular,  we observe that \OurModel{} significantly outperforms a comparable model that leverages only the visual modality and a comparable model that leverages the visual and attribute modality.

\keywords{Multimodal retrieval \and Category-to-image retrieval  \and E-commerce}
\end{abstract}

\section{Introduction}
\label{sec:intro}

Multimodal retrieval is a major but understudied problem in e-commerce~\cite{tsagkias-2020-challenges}. Even though e-commerce products are associated with rich multi-modal information,  research currently focuses mainly on textual and behavioral signals to support product search and recommendation. 
The majority of prior work in multimodal retrieval for e-commerce focuses on applications in the fashion domain, such as recommendation of fashion items~\cite{lin2019improving} and cross-modal fashion retrieval~\citep{laenen2019multimodal, goei2021tackling}.
In the more general e-commerce domain, multimodal retrieval has not been explored that well yet~\citep{hewawalpita2019multimodal,  li2020aspect}. 
The multimodal problem on which we focus is motivated by the importance of category information in e-commerce.  Product category trees are a key component of modern e-commerce as they assist customers when navigating across large and dynamic product catalogues~\citep{wirojwatanakul2019multi, tagliabue2020grow,kondylidis2021category}.  
Yet, the ability to retrieve an image for a given product category remains a challenging task mainly due to noisy category and product data, and the size and dynamic character of product catalogues~\citep{laenen2018web,tsagkias-2020-challenges}.

\header{The \acl{CtI} retrieval task}
We introduce the problem of retrieving a ranked list of relevant images of products that belong to a given category,  which we call the \acfi{CtI} retrieval task. 
Unlike image classification tasks that operate on a predefined set of classes,  in the \ac{CtI} retrieval task we want to be able not only to understand which images belong to a given category but also to generalize towards unseen categories.
Consider the category ``Home decor.'' 
A \ac{CtI} retrieval should output a ranked list of $k$ images retrieved from the collection of images that are relevant to the category, which could be anything from images of carpets to an image of a clock or an arrangement of decorative vases.
Use cases that motivate the \ac{CtI} retrieval task include 
\begin{enumerate*}[label=(\arabic*)]
\item the need to showcase different categories in search and recommendation results~\citep{tagliabue2020grow, tsagkias-2020-challenges, kondylidis2021category};
\item the task can be used to infer product categories in the cases when product categorical data is unavailable, noisy, or incomplete~\cite{yashima2016learning}; and
\item the design of cross-categorical promotions and product category landing pages~\cite{nielsen2000commerce}.
\end{enumerate*}

The \ac{CtI} retrieval task has several key characteristics:
\begin{enumerate*}[label=(\arabic*)]
\item we operate with categories from non-fixed e-commerce category trees, which range from very general (such as ``Automative'' or ``Home \& Kitchen'') to very specific ones (such as ``Helmet Liners'' or ``Dehumidifiers''). The category tree is not fixed, therefore, we should be able to generalize towards unseen categories; and 
\item product information is highly multimodal in nature; apart from category data, products may come with textual,  visual, and attribute information.
\end{enumerate*}

\header{A model for \ac{CtI} retrieval} To address the \ac{CtI} retrieval task, we propose a model that leverages image, text, and attribute information, \OurModel{}.  
\OurModel{} extends upon Contrastive Language-Image Pre-Training (CLIP)~\cite{radford2021learning}.  
\OurModel{} extends CLIP with the ability to represent attribute information.
Hence, \OurModel{} is able to use textual, visual, and attribute information for product representation. 
We compare the performance of \OurModel{} with several baselines such as unimodal BM25,  bimodal zero-shot CLIP, and 
MPNet~\cite{song2020mpnet}.  For our experiments, we use the XMarket dataset that contains textual, visual, and attribute information of e-commerce products~\citep{bonab21}.

\header{Research questions and contributions}
We address the following research questions:
\begin{enumerate*}[label=(RQ\arabic*)]
	\item How do baseline models perform on the \ac{CtI} retrieval task? Specifically, how do unimodal and bi-modal baseline models perform? How does the performance differ w.r.t.\ category granularity?
	\item How does a model, named CLIP-I, that uses product image information for building product representations impact the performance on the \ac{CtI} retrieval task?
	\item How does CLIP-IA, which extends CLIP-I with product attribute information, perform on the \ac{CtI} retrieval task?
	\item And finally, how does \OurModel{}, which extends CLIP-IA with product text information, perform on the \ac{CtI} task?
\end{enumerate*}

Our main contributions are:
\begin{enumerate*}[label=(\arabic*)]
	\item We introduce the novel task of \ac{CtI} retrieval and motivate it in terms of e-commerce applications.
	\item We propose \OurModel{}, the first model specifically designed for this task.  \OurModel{} leverages multimodal product data such as textual, visual, and attribute data. On average, \OurModel{} outperforms CLIP-I on all categories by 217\% and CLIP-IA by 269\%.  We share our code and experimental settings to facilitate reproducibility of our results.\footnote{\url{https://github.com/mariyahendriksen/ecir2022_category_to_image_retrieval}}
\end{enumerate*}

\section{Related Work}
\label{sec:related_work}

\textbf{Learning multimodal embeddings.}
Contrastive pre-training has been shown to be highly effective in learning joined embeddings across modalities~\cite{radford2021learning}. 
By predicting the correct pairing of image-text tuples in a batch, the CLIP model can learn strong text and image encoders that project to joint space. 
This approach to learning multimodal embeddings offers key advantages over approaches that use manually assigned labels as supervision: 
\begin{enumerate*}[label=(\arabic*)]
\item the training data can be collected without manual annotation; real-world data in which image-text pairs occur can be used;
\item models trained in this manner learn more general representations that allow for zero-shot prediction.
\end{enumerate*}
These advantages are appealing for e-commerce, as most public multimodal e-commerce datasets primarily focus on fashion only~\cite{bonab21}; being able to train from real-world data avoids the need for costly data annotation.

We build on CLIP by extending it to category-product pairs, taking advantage of its ability to perform zero-shot retrieval for a variety semantic concepts.

\header{Multimodal image retrieval} 
Early work in image retrieval grouped images into a restricted set of semantic categories and allowed users to retrieve images by using category labels as queries~\cite{smeulders2000}.  
Later work allowed for a wider variety of queries ranging from natural language~\citep{hu2016natural, vo2019composing}, to attributes~\cite{nagarajan2018attributes}, to combinations of multiple modalities (e.g., title, description, and tags)~\cite{thomee2016yfcc100m}.  Across these multimodal image retrieval approaches we find three common components:
\begin{enumerate*}[label=(\arabic*)]
\item an image encoder,
\item a query encoder, and 
\item a similarity function to match the query to images~\cite{radford2021learning,gupta2020contrastive}.
\end{enumerate*}
Depending on the focus of the work some components might be pre-trained, whereas the others are optimized for a specific task.  

In our work, we rely on pre-trained image and text encoders but learn a new multimodal composite of the query to perform \ac{CtI} retrieval.

\header{Multimodal retrieval in e-commerce} Prior work on multimodal retrieval in e-commerce has been mainly focused on cross-modal retrieval for fashion~\citep{zoghbi-2016-cross-modal, laenen2017cross, goei2021tackling}.  
Other related examples include outfit recommendation~\citep{lin2019improving, laenen-2020-comparative, li2020hierarchical}
Some prior work on interpretability for fashion product retrieval proposes to leverage multimodal signals to improve explainability of latent features~\citep{liao2018interpretable,  yang2019interpretable}.  
\citet{tautkute2019deepstyle} propose a multimodal search engine for fashion items and furniture.
When it comes to combining signals for improving product retrieval,  \citet{yim2018one} propose to combine product images,  titles,  categories, and descriptions to improve product search,  \citet{yamaura2019resale} propose an algorithm that leverages multimodal product information for predicting a resale price of a second-hand product.

Unlike prior work on multimodal retrieval in e-commerce that mainly focuses on fashion data,  we focus on creating multimodal product representations for the general e-commerce domain.

\section{Approach}
\label{sec:methodology}

\header{Task definition}
We follow the same notation as in~\citep{zhang2020contrastive}.
The input dataset can be presented as category-product pairs $(\bx_c, \bx_p)$, where $\bx_c$ represents a product category,  and $\bx_p$ represents information about product that belong to the category $\bx_c$.  
The product category $\bx_c$ is taken from the category tree $T$ and is represented as a category name.
The product information comprises titles $\bx_t$,  images $\bx_i$,  and attributes $\bx_i$,  i.e.,  $\bx_p = \{ \bx_i,  \bx_t, \bx_a  \}$.

For the \ac{CtI} retrieval task,  we use the target category name $\bx_c$ as a query and we aim to refturn a ranked list of top-$k$ images that belong to the category $\bx_c$.

\begin{figure}[t]
\centering
      \includegraphics[width=\textwidth]{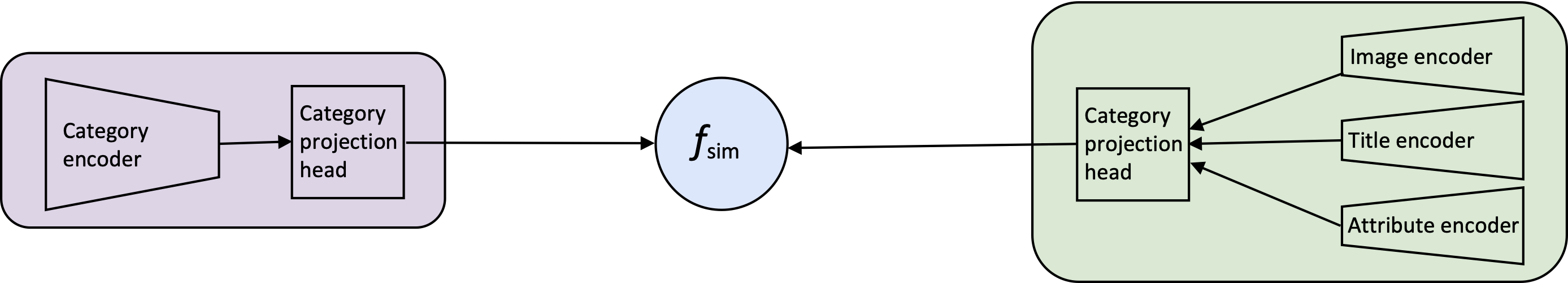}
      \caption{Overview of \OurModel{}. The category encoding pipeline is in purple;  the category information pipeline in green;
 $f_{sim}$ is a cosine similarity function.
}
  \label{fig:model}
\end{figure}

\header{\OurModel{}}
Fig.~\ref{fig:model} provides a high-level view of \OurModel{}.  \OurModel{} projects category $\bx_c$ and product information $\bx_p$ into a $d$-dimensional multimodal space where the resulting vectors are respectively $\bc$ and $\bp$.  The category and product information is processed by a category encoding pipeline and product information encoding pipeline. 
The core components of \OurModel{} are the encoding and projection modules.
The model consists out of four encoders: a category encoder,  an image encoder,  a title encoder,  and an attribute encoder.  Besides, \OurModel{} comprises two non-linear projection heads: the category projection head and the multimodal projection head.  

While several components of \OurModel{} are based on CLIP~\cite{radford2021learning}, \OurModel{} differs from CLIP in three important ways:
\begin{enumerate*}[label=(\arabic*)]
\item unlike CLIP, which operates on two encoders (textual and visual), \OurModel{} extends CLIP towards a category encoder,  image encoder,  textual encoder,  and attribute encoder;
\item \OurModel{} features two projection heads, one for the category encoding pipeline, and one for the product information encoding pipeline; and
\item while CLIP is trained on text-image pairs,  \OurModel{} is trained on category-product pairs, where product representation is multimodal.
\end{enumerate*}

\subheader{Category encoding pipeline}
The \emph{category encoder} ($f_c$) takes as input category name $\bx_c$ and returns its representation $\bh_c$. More specifically, we pass the category name  $\bx_c$ through the category encoder $f_c$:
\begin{equation}
\bh_c = f_c (\bx_c).
\end{equation}
To obtain this representation,  we use pre-trained MPNet model~\cite{song2020mpnet}. 
After passing category information through the category encoder, we feed it to the category projection head. 
The \emph{category projection head} ($g_c$) takes as input a query representation $\bh_c$  and projects it into $d$-dimensional multi-modal space:
\begin{equation}
\bc = g_c(\bh_c),
\end{equation}
where $\bc \in \BR^d$.

\subheader{Product encoding pipeline}
The product information encoding pipeline represents three encoders,  one for every modality,  and a product projection head.
The \emph{image encoder} ($f_i$) takes as input a product image $\bx_i$ aligned with the category $\bx_c$.  Similarly to the category processing pipeline,  we pass the product image $\bx_i$ through the image encoder:
\begin{equation}
\bh_i = f_i (\bx_i).
\end{equation}
To obtain the image representation $\bh_i$,  we use pre-trained Vision Transformer from CLIP model.
The \emph{title encoder} ($f_t$) takes a product title $\bx_t$ as input and returns a title representation $\bh_t$:
\begin{equation} 
\bh_t = f_t (\bx_t).
\end{equation}
Similarly to the category encoder $f_c$, we use pre-trained MPNet to obtain the title representation $\bh_t$. 
The \emph{attribute encoder} ($f_a$) is a network that takes as input a set of attributes $\bx_a = \{a_1, a_2,  \dots ,  a_n\} $ and returns their joint representation:
\begin{equation}
\bh_a = f_a(\bx_a) = \frac{1}{n} \sum_{i=1}^{n} f_a(\bx_{ai}).
\end{equation}
Similarly to the category encoder $f_c$ and title encoder $f_t$, we obtain representation of each attribute with the pre-trained MPNet model.
After obtaining title, image and attribute representations, we pass the representations into the product projection head.
The \emph{product projection head} ($g_p$) takes as input a concatenation of the image representation $\bh_i$, title representation $\bh_t$, and attribute representation $\bh_a$ and projects the resulting vector $\bh_p =  concat(\bh_i,  \bh_t,  \bh_a) $ into multimodal space:
\begin{equation}
\bp = g_p(\bh_p) = g_p(concat(\bh_i,  \bh_t,  \bh_a)),
\end{equation}
where $\bp \in \BR^d$.

\subheader{Loss function} 
We train \OurModel{} using bidirectional contrastive loss~\citep{zhang2020contrastive}.  The loss is a weighted combination of two losses: a category-to-product contrastive loss and a product-to-category contrastive loss.  In both cases the loss is the InfoNCE loss~\cite{oord2018representation}.  Unlike prior work that focuses on a contrastive loss between inputs of the same modality~\citep{he2020momentum, chen2020simple} and on corresponding inputs of two modalities~\cite{zhang2020contrastive},  we use the loss to work with inputs from textual modality (category representation) vs.\  a combination of multiple modalities (product representation).
We train \OurModel{} on batches of category-product pairs $(\bx_c, \bx_p)$ with batch size $\beta$.
For the $j$-th pair in the batch,  the category-to-product contrastive loss is computed as follows:
\begin{equation}
    \ell^{(c \rightarrow p)}_{j} = - \log \frac{ \exp( f_{sim}(\bc_j, \bp_j) / \tau) }{ \sum^{\beta}_{k=1} \exp ( f_{sim}(\bc_j, \bp_k) / \tau)},
\end{equation}
where $f_{sim}(\bc_i, \bp_i)$ is the cosine similarity,  and $\tau \in \R^+$ is a temperature parameter.
Similarly,  the product-to-category loss is computed as follows:
\begin{equation}
    \ell^{(p \rightarrow c)}_{j} = - \log \frac{ \exp( f_{sim}(\bp_j, \bc_j) / \tau) }{ \sum^{\beta}_{k=1} \exp ( f_{sim}(\bp_j, \bc_k) / \tau)}.
\end{equation}
The resulting contrastive loss is a combination of the two above-mentioned losses:
\begin{equation}
    \sL = \frac{1}{\beta} \sum^{\beta}_{j=1} \Big( \lambda \ell^{(p \rightarrow c)}_{j} + (1-\lambda) \ell^{(c \rightarrow p)}_{j} \Big),
\end{equation}
where $\beta$ represents the batch size and $\lambda \in [0,1]$ is a scalar weight.

\section{Experimental Setup}
\label{sec:experiments}

\shrink
\textbf{Dataset.}
\label{sec:dataset}
We use the XMarket dataset recently introduced by~\citet{bonab21}  that contains textual,  visual,  and attribute information of e-commerce products as well as a category tree. For our experiments, we select \numprint{38921} products from the US market.
Category information is represented as a category tree and comprises \numprint{5471} unique categories across nine levels.  Level one is the most general category level, level nine is the most specific level.  Every product belongs to a subtree of categories $t \in T$.  In every subtree $t$, each parent category has only one associated child category.
The average subtree depth is \numprint{4.63} (minimum: 2, maximum: 9).
Because every product belongs to a subtree of categories,  the dataset contains \numprint{180094} product-category pairs in total.
We use product titles as textual information and one image per product as visual information.
The attribute information comprises  \numprint{228368} attributes, with \numprint{157049} unique.  On average,  every product has  \numprint{5.87} attributes (minimum: 1, maximum: 24).

\header{Evaluation method}
To investigate how model performance changes w.r.t. category granularity, for every product in the dataset, $\bx_p$,  and the corresponding subtree of categories to which the product belongs,  $t$,  we train and evaluate the model performance in three settings:
\begin{enumerate*}[label=(\arabic*)]
	\item \textit{all categories}, where we randomly select one category from the subtree $t$;
	\item \textit{most general category}, where we use only the most general category of the subtree $t$, i.e.,  the root; and 
	\item \textit{most specific category}, where we use the most specific category of the subtree $t$.
\end{enumerate*}
In total, there are \numprint{5471} categories in all categories setup, 34 categories in the most general category, and \numprint{4100} in the most specific category setup.
We evaluate every model on category-product pairs $(\bx_c, \bx_p)$ from the test set. We encode each category and a candidate product data by passing them through category encoding and product information encoding pipelines. For every category $\bx_c$ we retrieve the top-$k$ candidates ranked by cosine similarity w.r.t.\ the target category $\bx_c$.

\header{Metrics} 
To evaluate model performance,  we use Precision@K where $K = \{1, 5, 10 \}$,   mAP@K where $K = \{ 5,  10 \}$,  and R-precision.

\header{Baselines}
Following~\citep{wang2021bert, shen2021much, dai2020funnel} we use BM25,  MPNet,  CLIP as our baselines.

\header{Four experiments}
We run four experiments, corresponding to our research questions as listed at the end of Section~\ref{sec:intro}. 
In \emph{Experiment~1} we evaluate the baselines on the \ac{CtI} retrieval task (RQ1).
We feed BM25 corpora that contain textual product information, i.e., product titles. We use MPNet in a zero-shot manner. For all the products in the dataset, we pass the product title $\bx_t$ through the model. During the evaluation, we pass a category  $\bx_c$ expressed as textual query through MPNet and retrieve top-$k$ candidates ranked by cosine similarity w.r.t.\ the target category $\bx_c$. We compare categories of the top-$k$ retrieved candidates with the target category $\bx_c$. 
Besides, we use pre-trained CLIP in a zero-shot manner with a Text Transformer and a Vision Transformer (ViT)~\cite{dosovitskiy2021} an configuration. We pass the product images $\bx_i$ through the image encoder.
For evaluation, we pass a category  $\bx_c$ through the text encoder and retrieve top-$k$ image candidates ranked by cosine similarity w.r.t.\ the target category $\bx_c$. We compare categories of the top-$k$ retrieved images with the target category $\bx_c$. 

In \emph{Experiment~2} we evaluate image-based product representations (RQ2). After obtaining results with CLIP in a zero-shot setting, we build product representations by training on e-commerce data.  First, we investigate how using product image data for building product representations impacts performance on the \ac{CtI} retrieval task. 
To introduce visual information, we extend CLIP in two ways:
\begin{enumerate*}[label=(\arabic*)]
\item We use ViT from CLIP as image encoder $f_i$. We add product projection head $g_p$ that takes as an input product visual information $\bx_i \in \bx_p$. 
\item We use the text encoder from MPNet as category encoder $f_c$; we add a category projection head $g_c$ on top of category encoder $f_c$ thereby completing category encoding pipeline (see Fig.~\ref{fig:model}).
\end{enumerate*}
We name the resulting model CLIP-I.
We train CLIP-I on category-product pairs $(\bx_c, \bx_p)$ from the training set. Note that $\bx_p = \{ \bx_i \}$, i.e.,  we only use visual information for building product representations.

In \emph{Experiment~3}, we evaluate image- and attribute-based product representations (RQ3). 
We extend CLIP-I by introducing attribute information to the product information encoding pipeline.  
We add an attribute encoder $f_a$ through which we obtain a representation of product attributes,  $\bh_a$.  We concatenate the resulting attribute representation with image representation $\bh_p =  concat(\bh_i,  \bh_a)$ and pass the resulting vector to the product projection head $g_p$. Thus, the resulting product representation $\bp$ is based on both visual and attribute product information. 
We name the resulting model CLIP-IA.
We train CLIP-IA on category-product pairs $(\bx_c, \bx_p)$ where $\bx_p = \{ \bx_i, \bx_a \}$,  i.e.,  we use visual and attribute information for building product representation.

In \emph{Experiment~4}, we evaluate image- attribute-, and title-based product representations (RQ4). 
We investigate how extending the product information processing pipeline with the textual modality impacts performance on the \ac{CtI} retrieval task. 
We add title encoder $f_t$ to the product information processing pipeline and use it to obtain title representation $\bh_t$.  
We concatenate the resulting representation with product image and attribute representations $\bh_p =  concat(\bh_i,  \bh_t,  \bh_a)$. We pass the resulting vector to the product projection head $g_p$.
The resulting model is \OurModel{}.
We train and test \OurModel{} on category-product pairs $(\bx_c, \bx_p)$ where $\bx_p = \{ \bx_i, \bx_a, \bx_t \}$,  i.e.,  we use visual,  attribute, and textual information for building product representations.

\header{Implementation details}
We train every model for 30 epochs, with a batch size $\beta = 8$ for most general categories,  $\beta = 128$ --- for most specific categories and all categories.  For loss function, we set $\tau = 1$,  $\lambda = 0.5$.  We implement every projection head as non-linear MLPs with two hidden layers,  GELU non-linearities~\citep{hendrycks2016gaussian} and layer normalization~\citep{ba2016layer}. We optimize both heads with the AdamW optimizer~\cite{loshchilov2017decoupled}.

\section{Experimental results}

\begin{table}[t]
\centering
\caption{Results of Experiments 1--4. The best performance is highligthed in bold.}
\label{tab:experiments_results}
\begin{tabular}{lrrrrrr}
\toprule
\textbf{Model} & \multicolumn{1}{c}{\textbf{P@1}} & \multicolumn{1}{c}{\textbf{P@5}} & \multicolumn{1}{c}{\textbf{P@10}} & \multicolumn{1}{c}{\textbf{MAP@5}} & \multicolumn{1}{c}{\textbf{MAP@10}} & \multicolumn{1}{c}{\textbf{R-precision}} \\
\cmidrule(l){2-4} \cmidrule(l){5-6} \cmidrule(l){7-7} 
\multicolumn{7}{c}{\textbf{All categories (\numprint{5471}) } }                                                                                                                                                                                                                                      \\
\midrule
BM25~\cite{jones2000probabilistic}        & \numprint{0.01} & \numprint{0.01}  & \numprint{0.01}  & \numprint{0.01}  & \numprint{0.01}  & \numprint{0.01}  \\
CLIP~\cite{radford2021learning}  & \numprint{0.01} & \numprint{0.02}  & \numprint{0.02}  & \numprint{0.03}  & \numprint{0.04}  & \numprint{0.02}  \\
MPNet~\cite{song2020mpnet}  & \numprint{0.01} & \numprint{0.06}  & \numprint{0.06}  & \numprint{0.07}  & \numprint{0.09}  & \numprint{0.05}  \\

CLIP-I  (Ours)            & \numprint{3.3}  & \numprint{3.8}   & \numprint{3.79}  & \numprint{6.81}  & \numprint{7.25}  & \numprint{3.67}  \\
CLIP-IA  (Ours)          & \numprint{2.5}  & \numprint{3.34}  & \numprint{3.29}  & \numprint{5.95}  & \numprint{6.24}  & \numprint{3.27}  \\
CLIP-ITA (Ours)        & \textbf{\numprint{9.9}}  &\textbf{ \numprint{13.27}} & \textbf{\numprint{13.43}} & \textbf{\numprint{20.3}}  &\textbf{\numprint{20.53}} & \textbf{\numprint{13.42}}                                            \\
\midrule
\multicolumn{7}{c}{\textbf{Most general category  (\numprint{34})}}                                                                                                                                                                                                                   \\
\midrule
BM25~\cite{jones2000probabilistic}                & \numprint{2.94}           & \numprint{4.71}           & \numprint{4.71}           & \numprint{8.33}           & \numprint{8.28}           & \numprint{4.48}           \\
CLIP~\cite{radford2021learning}   & \numprint{11.76}          & \numprint{12.35}          & \numprint{11.76}          & \numprint{16.12}          & \numprint{15.18}          & \numprint{9.47}           \\
MPNet~\cite{song2020mpnet}   & \numprint{14.70}           & \numprint{15.8}           & \numprint{15.01}          & \numprint{18.44}          & \numprint{18.78}          & \numprint{9.35}           \\

CLIP-I   (Ours)          & \numprint{17.85}          & \numprint{17.14}          & \numprint{16.78}          & \numprint{19.88}          & \numprint{20.14}          & \numprint{13.02}          \\
CLIP-IA    (Ours)        & \numprint{21.42}          & \numprint{21.91}          & \numprint{22.78}          & \numprint{25.59}          & \numprint{26.29}          & \numprint{20.74}          \\
CLIP-ITA  (Ours)         & \textbf{\numprint{35.71}} &\textbf{ \numprint{30.95}} & \textbf{ \numprint{30.95}} & \textbf{\numprint{35.51}} & \textbf{\numprint{34.28}} &\textbf{ \numprint{25.79}} \\
\midrule
\multicolumn{7}{c}{\textbf{Most specific category (\numprint{4100}) }}                                                                                                                                                                                                              \\
\midrule
BM25~\cite{jones2000probabilistic}                & \numprint{0.02}  & \numprint{0.02}  & \numprint{0.01}  & \numprint{0.01}  & \numprint{0.01}  & \numprint{0.01}  \\
CLIP~\cite{radford2021learning}   & \numprint{11.92} & \numprint{9.81}  & \numprint{9.23}  & \numprint{15.12} & \numprint{14.95} & \numprint{8.14}  \\
MPNet~\cite{song2020mpnet}   & \numprint{33.36} & \numprint{28.56} & \numprint{26.93} & \numprint{37.43} & \numprint{36.77} & \numprint{25.29} \\

CLIP-I   (Ours)          & \numprint{14.06} & \numprint{12.11} & \numprint{11.53} & \numprint{18.24} & \numprint{17.9}  & \numprint{11.22} \\
CLIP-IA  (Ours)          & \numprint{35.3}  & \numprint{30.21} & \numprint{29.32} & \numprint{39.93} & \numprint{39.27} & \numprint{28.86} \\
CLIP-ITA  (Ours)         & \textbf{\numprint{45.85}} &  \textbf{\numprint{41.04}} & \textbf{\numprint{40.02}} & \textbf{\numprint{50.04}} & \textbf{\numprint{49.87}} & \textbf{\numprint{39.69}}                                    \\
\bottomrule   
\end{tabular}
\end{table}

\shrink
\noindent%
\textbf{Experiment 1: Baselines.}
Following RQ1, we start by investigating how do baselines perform on \ourtask{}. Besides, we investigate how does the performance on the task differs between the unimodal and the bimodal  approach.

The results are shown in Table~\ref{tab:experiments_results}. 
When evaluating on all categories,  all the baselines perform poorly.
For the most general category setting, MPNet outperforms CLIP on all metrics except R-precision.  The most prominent gain is for Precision@10 where MPNet outperforms CLIP by 28\%. CLIP outperforms BM25 on all metrics.
For the most specific category setting, MPNet performance is the highest, BM25 --- the lowest. In particular, MPNet outperforms CLIP by 211\% in Precision@10.
Overall,  MPNet outperforms CLIP and both models significantly outperforms BM25 for both most general and most specific categories. However, when evaluation is done on all categories, the performance of all models is comparable.
As an answer to RQ1, the results suggest that using information from multiple modalities is beneficial for performance on the task.

\header{Experiment 2: Image-based product representations}
To address RQ2, we compare the performance of CLIP-I with CLIP and MPNet, the best-performing baseline.
Table~\ref{tab:experiments_results},  shows the experimental results for Experiment 2.
The biggest performance gains are obtained in ``all categories'' setting. However, there, the performance of the baselines was very poor.
For the most general categories,  CLIP-I outperforms both CLIP and MPNet.  For CLIP-I vs. CLIP, we observe the biggest increase of 51\% for Precision@1, for CLIP-I vs. MPNet --- 39\% in R-precision.
In the case of the most specific categories,  CLIP-I outperforms CLIP but loses to MPNet.
Overall, CLIP-I outperforms CLIP in all three settings and outperforms MPNet except the most specific categories.  Therefore, we answer RQ2 as follows: the results suggest that extension of CLIP by the introduction of product image data for building product representations has a positive impact on performance on \ourtask{}.

\header{Experment 3: Image- and attribute-based product representations}
To answer RQ3, we compare the performance of CLIP-IA with CLIP-I and the baselines.
The results are shown in Table~\ref{tab:experiments_results}. 
When evaluated on all categories, CLIP-IA performs worse than CLIP-I but outperforms MPNet.  In particular, CLIP-I obtains the biggest gain relative of 32\% on Precision@1 and the lowest gain of 12\% on R-precision.
For the most general category,  CLIP-IA outperforms CLIP-I and MPNet on all metrics.  More specifically,  we observe the biggest gain of 122\% on R-precision over MPNet and the biggest gain of 59\% on R-precision for CLIP-I. 
Similarly,  for the most specific category,  CLIP-IA outperforms both CLIP-I and MPNet. We observe the biggest relative gain of 138\% over CLIP-I.
The results suggest that further extension of CLIP by the introduction of the product image and attribute data for building product representations has a positive impact on performance on \ourtask{}, especially when evaluated on most specific categories. Therefore, we answer RQ4 positively.

\header{Experiment 4: Image-, attribute-, and title-based product representations}
We compare \OurModel{} with both CLIP-IA, CLIP-I, and the baselines. The results are shown in Table~\ref{tab:experiments_results}. 
In general, \OurModel{} outperforms CLIP-I and CLIP-IA and the baselines in all settings.
When evaluated on all categories,  the maximum relative increase of CLIP-ITA over CLIP-I is 265\% in R-precision,  the minimum relative increase is 183\% in mAP@10.
The biggest relative increase of CLIP-ITA performance over CLIP-IA is 310\% in Precision@1,  the smallest relative increase is 229\% in mAP@10.
For the most general categories,  \OurModel{} outperforms CLIP-I by 82\% and CLIP-IA by 38\%.
For most specific categories, we observe the biggest increase of CLIP-ITA over CLIP-I of 254\% in R-precision and the smallest relative increase of 172\% on mAP@5.
At the same time,  the biggest relative increase of CLIP-ITA over CLIP-IA is a 38\% increase in R-precision and the smallest relative increase is a 27\% increase in mAP@5.
Overall,  CLIP-ITA wins in all three settings.  Hence,  we answer RQ4 positively.

\section{Error Analysis}
\label{sec:analysis}

\shrink
\textbf{Distance between predicted and target categories.}
We examine the performance of \OurModel{} by looking at the pairs of the ground-truth and predicted categories $(c,  c_p)$ in cases when the model failed to predict the correct category, i.e.,  $c \neq c_p$. This allows us to quantify how far off the incorrect predictions lie w.r.t.\ the category tree hierarchy.
First, we examine in how many cases target category $c$ and predicted category $c_p$  belong to the same most general category, i.e., belong to the same category tree; see Table~\ref{tab:same_tree_vs_different_tree}. 
In the case of most general categories,  the majority of incorrectly predicted categories belong to a tree different from the target category tree.  For the most specific categories,  about 11\% of predicted categories belong to the category tree of the target category.  However,  when evaluation is done on all categories,  72\% of incorrectly predicted cases belong to the same tree as a target category.

\begin{table}[t]
\centering
\caption{Erroneous \OurModel{} prediction counts for ``same tree'' vs. `` different tree'' predictions  per evaluation type.}
\label{tab:same_tree_vs_different_tree}
\begin{tabular}{lrr}
\toprule
                                           & \multicolumn{1}{l}{\textbf{Same tree}} & \multicolumn{1}{l}{\textbf{Different tree}} \\
\midrule
All categories                             & \numprint{1655}                                 & \numprint{639}                                   \\
The most general category   & \numprint{2}                                    & \numprint{21}                                    \\
The most specific category & \numprint{127}                                  & \numprint{1011}                                  \\
\midrule
Total                                      & \numprint{1786}                                 & \numprint{1671}                                  \\
\bottomrule 
\end{tabular}
\end{table}

\begin{figure}[t]
    \centering
    \begin{subfigure}{0.31\textwidth}
        \includegraphics[width=\textwidth]{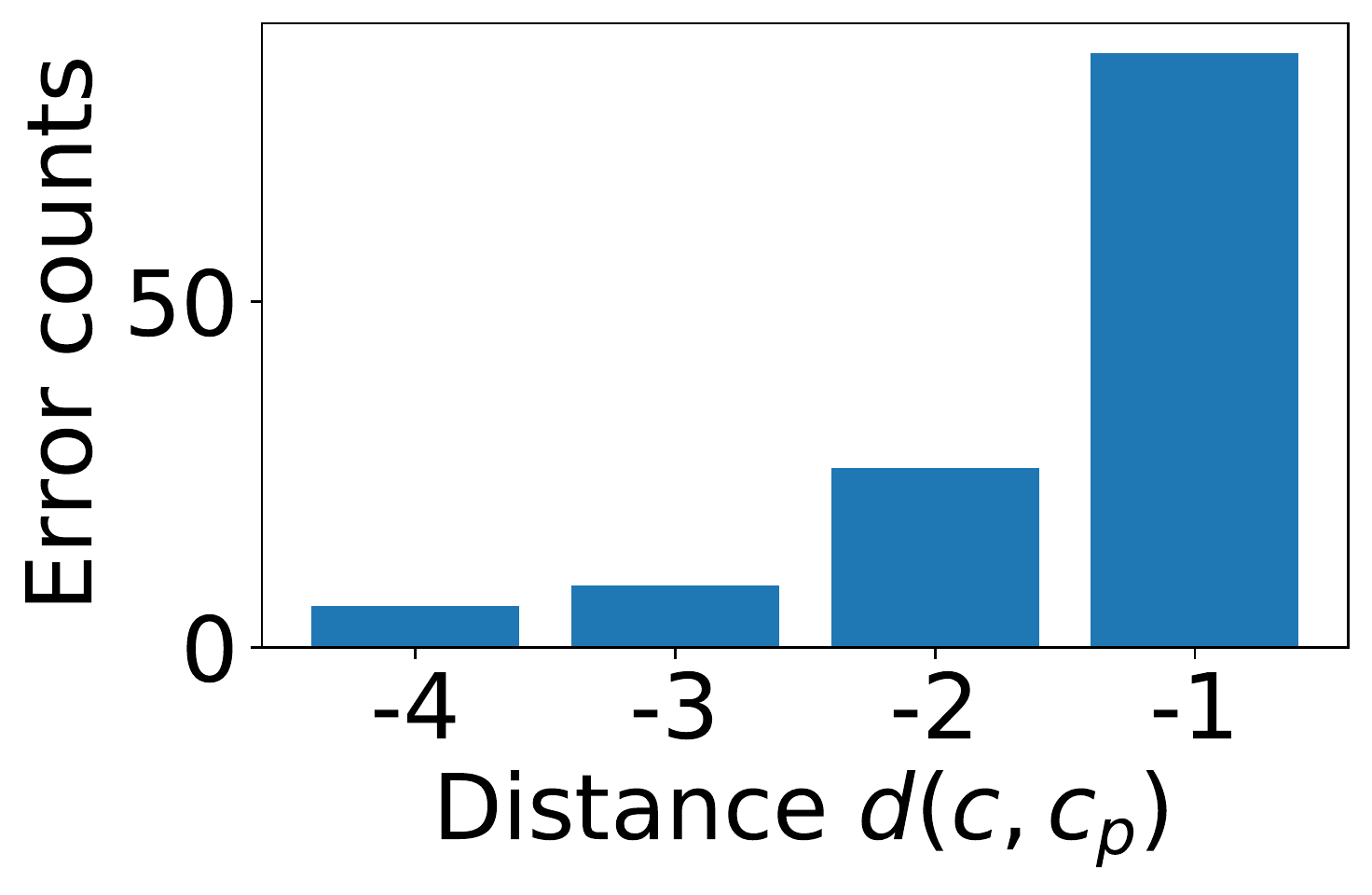}
        \vspace*{-6mm} 
        \caption{Most specific categories}
        \label{fig:clip_ita_most_spec_cat}
    \end{subfigure}
    \hfill
    \begin{subfigure}{0.31\textwidth}
        \includegraphics[width=\textwidth]{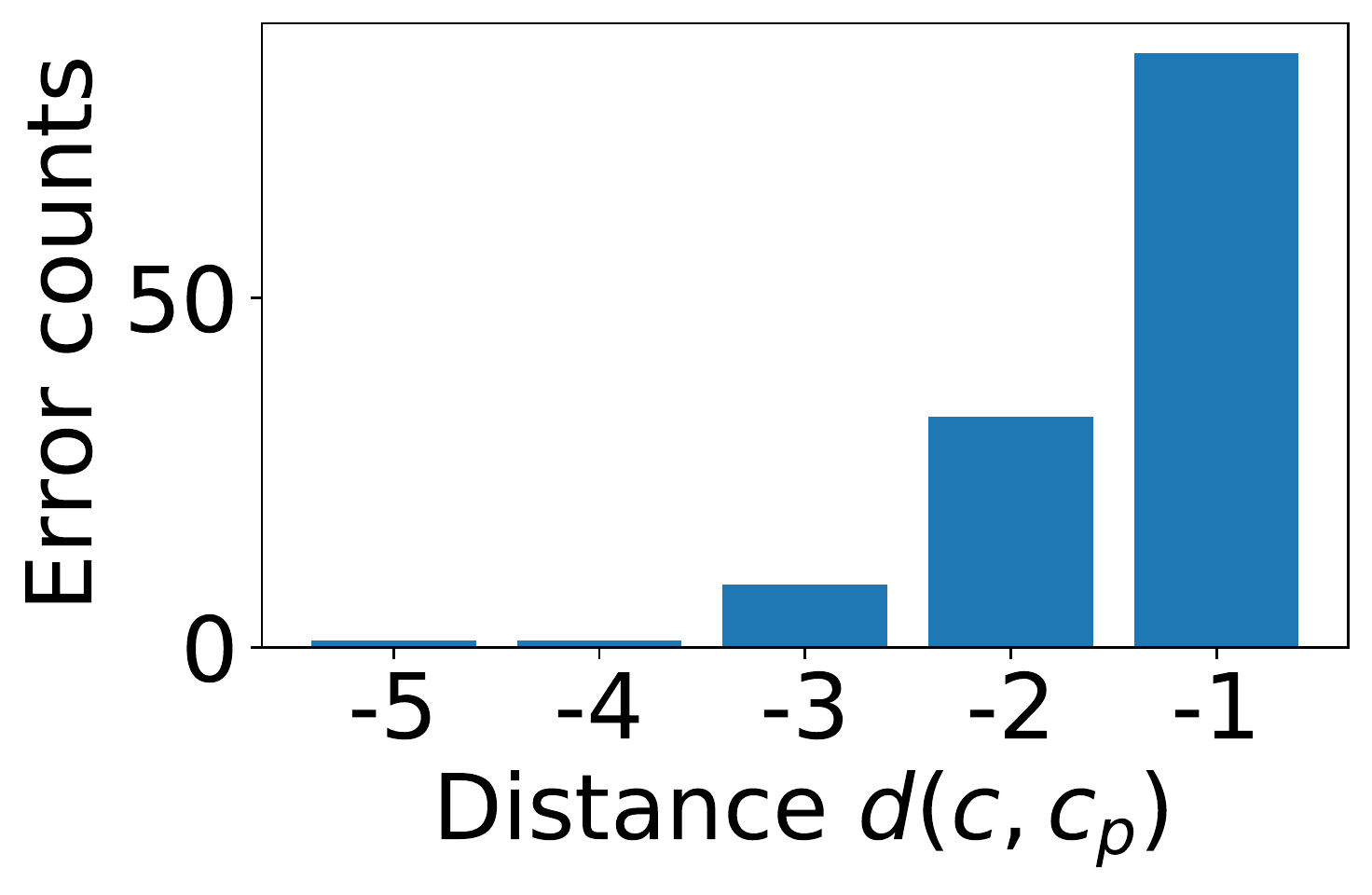}
        \vspace*{-6mm}
        \caption{All categories, $d < 0$}
        \label{fig:clip_ita_most_all_cat_left}
    \end{subfigure}
    \hfill
    \begin{subfigure}{0.31\textwidth}
        \includegraphics[width=\textwidth]{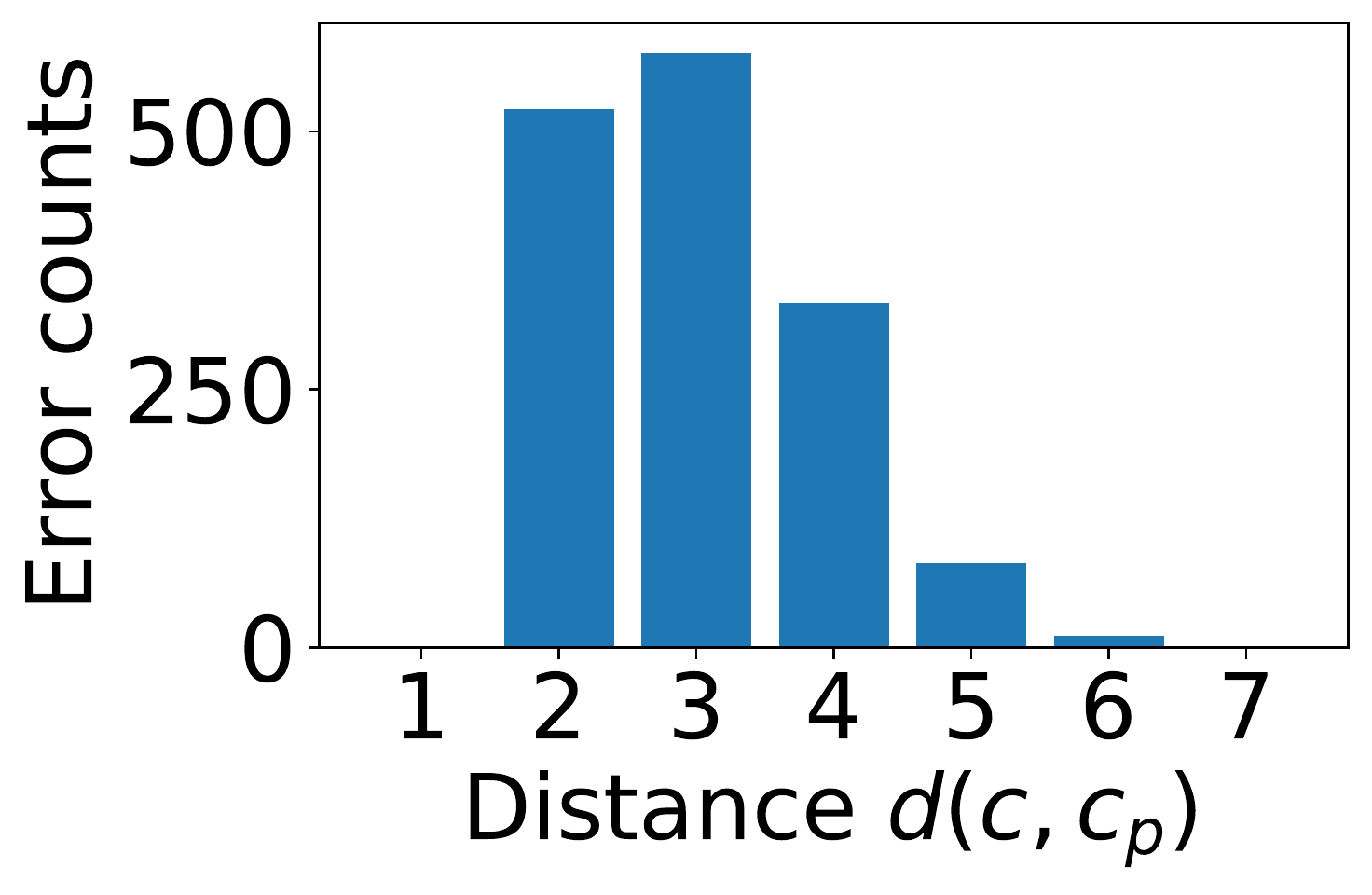}
        \vspace*{-6mm}
        \caption{All categories,  $d > 0$}
        \label{fig:clip_ita_most_all_cat_right}
    \end{subfigure}
    \caption{Error analysis for \OurModel{}. Distance between target category $c$ and a predicted category $c_p$ when $c$ and $c_p$ are in the same tree.}
    \label{fig:categorical_tree_distance}
\end{figure}

Next,  we turn to the category-predicted category pairs $(c,  c_p)$ where the incorrectly predicted category $c_p$ belongs to the same tree as target category $c$.  We compute the distance $d$ between a category used as a query $c$ and a predicted category $c_p$. 
We compute the distance between target category $c$ and a top-$1$ predicted category $c_p$ as the difference between their respective depths $d(c, c_p) = depth(c_p) - depth(c)$.  The distance $d$ is positive if the depth of the predicted category is bigger than the depth of the target category,  $depth(c_p) > depth(c)$,  i.e.,  the predicted category is more specific than the target category.  
The setup is mirrored for negative distances.
See Fig.~\ref{fig:categorical_tree_distance}. 
We do not plot the results for the most general category because for this setting there are only two cases when target category $c$ and a predicted category $c_p$ were in the same tree.  In both cases, predicted category $c_p$ was more general than target category $c$ with distance $d(c, p_c) = 2$. 
In cases when target category $c$ was sampled from the most specific categories,  the wrongly predicted category $c_p$ belonging to the same tree was always more specific than the target category $c$ with the maximum absolute distance between $c$ and $c_p$,  $|d(c, c_p)| = 4$.  In 68\% of the cases the predicted category was one level above the target category,  for 21\% $d(c, c_p) = -2$,  for 7\% $d(c, c_p) = -3$, and for 5\% $d(c, c_p) = -4$.
For the setting with all categories,  in 92\% of the cases, the predicted category $c_p$ was more specific than the target category $c$; for 8\% the predicted category was more general. 

Overall,  for the most general category and the most specific category,  the majority of incorrectly predicted categories are located in a category tree different from the one where the target category was located.  For the ``all categories'' setting, it is the other way around.
When it comes to the cases when incorrectly predicted categories are in the same tree as a target category,  the majority of incorrect predictions are 1 level more general when the target category is sampled from the most specific categories.  For the ``all categories'' setting,  the majority of incorrect predictions belonging to the same tree as the target category were more specific than the target category.
Our analysis suggests that efforts to improve the performance of \OurModel{} should focus on minimizing the (tree-based) distance between the target and predicted category in a category tree. This could be incorporated as a suitable extension of the loss function.

\header{Performance on seen vs.  unseen categories}
Next, we investigate how well \OurModel{} generalizes to unseen categories.  We split the evaluation results into two groups based on whether the category used as a query was seen during training or not; see Table~\ref{tab:seen_vs_unseen_cats}.
For the most general categories,  \OurModel{} is unable to correctly retrieve an image of the product of the category that was not seen during training at all.
For the most specific categories,  \OurModel{} performs better on seen categories than on unseen categories.  We observe the biggest relative performance increase of 85\% in mAP@10 and the smallest relative increase of 57\% in R-precision.
When evaluating on all categories,  \OurModel{} performs on unseen categories better when evaluated on Precision@k (27\% higher in Precision@1,  33\% higher in Precision@5, 10\% increase in Precision@10) and R-precision (relative increase of 32\%).  Performance on seen categories is better in terms of mAP@k (10\% increase for both mAP@5 and mAP@10).

Overall,  for the most general and most specific categories, the model performs much better on categories seen during training.  For ``all categories'' setting,  however,  \OurModel{}'s performance on unseen categories is better.

\begin{table}[t]
\centering
\caption{CLIP-ITA performance on seen vs.  unseen categories.}
\label{tab:seen_vs_unseen_cats}
\begin{tabular}{lrrrrrr}
\toprule
\textbf{Model} & \multicolumn{1}{c}{\textbf{P@1}} & \multicolumn{1}{c}{\textbf{P@5}} & \multicolumn{1}{c}{\textbf{P@10}} & \multicolumn{1}{c}{\textbf{mAP@5}} & \multicolumn{1}{c}{\textbf{mAP@10}} & \multicolumn{1}{c}{\textbf{R-precision}} \\

\cmidrule{2-4} \cmidrule(l){5-6} \cmidrule(l){7-7} 
         & \multicolumn{6}{c}{\textbf{All categories (\numprint{5471}) } }                                                               \\
\midrule 

CLIP-ITA (unseen cat.) & \numprint{13.3}  & \numprint{18.56} & \numprint{15.55} & \numprint{19.7}  & \numprint{19.65} & \numprint{18.52}       \\
CLIP-ITA (seen cat.)   & \numprint{10.48} & \numprint{13.95} & \numprint{14.08} & \numprint{21.65} & \numprint{21.65} & \numprint{14.07}       \\

\midrule
          & \multicolumn{6}{c}{\textbf{Most general category (\numprint{34})}}                                            \\
\midrule
CLIP-ITA (unseen cat.) & \numprint{0.0}    & \numprint{0.0}     & \numprint{0.0}    & \numprint{0.0}     & \numprint{0.0}    & \numprint{0.0}           \\
CLIP-ITA (seen cat.)   & \numprint{19.23} & \numprint{20.01} & \numprint{17.31} & \numprint{20.41} & \numprint{20.01} & \numprint{15.73}       \\
\midrule
         & \multicolumn{6}{c}{\textbf{Most specific category (\numprint{4100})}}                                             \\
\midrule
CLIP-ITA (unseen cat.) & \numprint{27.27} & \numprint{26.44} & \numprint{26.44} & \numprint{27.92} & \numprint{27.92} & \numprint{26.45}       \\
CLIP-ITA (seen cat.)   & \numprint{47.83} & \numprint{43.09} & \numprint{42.14} & \numprint{52.41} & \numprint{51.89} & \numprint{41.58}       \\

\bottomrule   
\end{tabular}
\end{table}

\section{Conclusion}
\label{sec:conclusion}

\shrink
We introduced the task of category-to-image retrieval and motivated its importance in the e-commerce scenario.  In the \ac{CtI} retrieval task,  we aim to retrieve an image of a product that belongs to the target category. 
We proposed a model specifically designed for this task, \OurModel{}.  \OurModel{} extends CLIP, one of the best performing text-image retrieval models. \OurModel{} leverages multimodal product data such as textual,  visual,  and attribute data to build product representations.
In our experiments, we contrasted and evaluated different combinations of signals from modalities, using three settings: on all categories, the most general, and the most specific categories.

We found that combining information from multiple modalities to build product representation produces the best results on the \ourtask{}.  \OurModel{} gives the best performance both on all categories and on the most specific categories. 
On the most general categories, CLIP-I, a model where product representation is based on image only, works slightly better.  CLIP-I performs worse on the most specific categories and across all categories.  For identification of the most general categories, visual information is more relevant.
Besides,  \OurModel{} is able to generalize to unseen categories except in the case of most general categories. However, the performance on unseen categories is lower than the performance on seen categories.
Even though our work is focused on the e-commerce domain, the findings can be useful for other areas, e.g., digital humanities. 

Limitations of our work are due to type of data in the e-commerce domain. In e-commerce, there is typically one object per image and the background is homogeneous, textual information is lengthly and noisy; in the general domain, there is typically more than one object per image, image captions are more informative and shorter.
Future work directions can focus on improving the model architecture. It would be interesting to incorporate attention mechanisms into the attribute encoder and explore how it influences performance. Another interesting direction for future work is to evaluate \OurModel{} on other datasets outside of the e-commerce domain. Future work can also focus on minimizing the distance between the target and predicted category in the category tree.

\header{Acknowledgements}
This research was supported by Ahold Delhaize, the Nationale Politie,  and the Hybrid Intelligence Center, a 10-year program funded by the Dutch Ministry of Education, Culture and Science through the Netherlands Organisation for Scientific Research, \url{https://hybrid-intelligence-centre.nl}.

All content represents the opinion of the authors, which is not necessarily shared or endorsed by their respective employers and/or sponsors.

\bibliographystyle{spbasic}
\bibliography{bibliography}

\end{document}